\documentclass[preprint,prd,noshowpacs,nofootinbib]{revtex4}
\usepackage{amssymb}
\usepackage{amsmath}
\usepackage{graphicx}

\begin{document}

\title{Time-dependent non-Abelian Aharonov-Bohm effect}

\author{Max Bright}
\email{neomaxprime@mail.fresnostate.edu}
\affiliation{Department of Physics, California State University Fresno, Fresno, California 93740-8031, USA}

\author{Douglas Singleton}
\email{dougs@csufresno.edu}
\affiliation{Department of Physics, California State University Fresno, Fresno, California 93740-8031, USA}

\date{\today}

\begin{abstract}
In this article, we study the {\it time-dependent} Aharonov-Bohm effect for non-Abelian gauge fields. We use two
well known time-dependent solutions to the Yang-Mills field equations to investigate the Aharonov-Bohm phase shift. For
both of the solutions, we find a cancellation between the phase shift coming from the non-Abelian ``magnetic'' field
and the phase shift coming from the non-Abelian ``electric'' field, which inevitably arises in time-dependent cases. 
We compare and contrast this cancellation for the time-dependent non-Abelian case to a similar cancellation which 
occurs in the time-dependent Abelian case. We postulate that this cancellation occurs generally in time-dependent
situations for both Abelian and non-Abelian fields. 
\end{abstract}

\maketitle
\section{Introduction}
The Aharonov-Bohm effect \cite{AB, ES} is usually investigated in terms of Abelian gauge theories, {\it e.g.} electromagnetism  
formulated via Maxwell's equations. Further, the electromagnetic fields considered in the canonical  Aharonov-Bohm effect 
are static fields. For the vector/magnetic Aharonov-Bohm  effect, this means a static vector potential, ${\bf A} ({\bf r})$,
which then translates to a static magnetic field via ${\bf B} = \nabla \times {\bf A}$. In this article, we
wish to consider the Aharonov-Bohm  effect in the presence of {\it time-dependent, non-Abelian} gauge fields. There has been 
some prior work on the Aharonov-Bohm effect in the presence of time-independent, non-Abelian fields \cite{Horvathy}. 
Unlike the Abelian case of electromagnetism, it may not be possible to observe the Aharonov-Bohm effect for static, non-Abelian 
fields. For the strong interaction, with the non-Abelian $SU(3)$ gauge group, the theory is thought to exhibit confinement. Thus,
it is not clear that one could arrange a non-Abelian flux tube that one could control, as is the case with electromagnetism. Further, 
since the color charges are always confined, one can not send isolated, unconfined color charges around hypothetical 
non-Abelian magnetic flux tubes, unlike the Abelian case of electromagnetism, where one can send isolated, unconfined electric 
charges around Abelian magnetic flux tubes. Despite these experimental obstacles, 
in this paper, we study the {\it time-dependent} Aharonov-Bohm effect for non-Abelian fields.
The first reason is that the Aharonov-Bohm effect is an important consequence of combining gauge theories with quantum
mechanics, and so, it is of interest to see how replacing an Abelian gauge theory by a non-Abelian gauge theory changes
(if at all) the Aharonov-Bohm effect. Second, the time-dependent Aharonov-Bohm effect has not been investigated to any great
degree, even for Abelian gauge theories. In the two papers \cite{singleton, singleton2}, the time-dependent Aharonov-Bohm effect
for Abelian fields was investigated and a cancellation was found between the usual magnetic Aharonov-Bohm phase shift and
the additional phase shift coming from the electric field, which inevitably occurs for time varying magnetic fields. In this
paper, we want to see if a similar cancellation occurs between the non-Abelian magnetic and electric fields.

For our time-dependent non-Abelian field configurations, we take the non-Abelian plane wave solutions of Coleman \cite{coleman} and 
the time-dependent Wu-Yang monopole solution \cite{arodz}. Both solutions satisfy the Yang-Mills field equations for
non-Abelian gauge fields of the form
\begin{equation}
\label{ym-field-eqn}
\partial ^\mu F^a _{\mu \nu} + g f^{abc} A^{\mu b} F^c _{\mu \nu} = 0 ~,
\end{equation}
where $g$ is the coupling constant and $f^{abc}$ are the group structure constants. $A^{\mu a}$ is the non-Abelian
vector potential and the field strength tensor, $F^a _{\mu \nu}$ is given by
\begin{equation}
\label{field-tensor}
F_{\mu \nu}^a = \partial _\mu A_\nu ^a - \partial _\nu A_\mu ^a + g f^{abc} A^b _\mu A^c _\nu ~.
\end{equation}
At this point in this paper, we will set $g=1$. For the Coleman non-Abelian plane wave solutions, we find 
the same cancellation between the non-Abelian magnetic and electric phase shifts that occur in the Abelian case. We also find the
same cancellation for the time-dependent Wu-Yang monopole solution. We conclude by giving some remarks as to the similarities
between these two time-dependent non-Abelian solutions and the time-dependent Abelian case. We further postulate
that the cancellation between the non-Abelian magnetic and electric phase shifts found for the two specific
solutions investigated here may be a feature of more general time-dependent non-Abelian solutions.   

\section{Time-dependent non-Abelian plane wave solution}
\label{plane-wave}
We begin by reviewing the properties of the two Coleman plane wave solutions. The non-Abelian
vector potential for the first/$(+)$ solution is 
\begin{equation}
\label{old-coleman-a}
A^{(+) a}_\mu = \left (x f^a(\zeta^{+}) + y g^a (\zeta ^{+}) + h^a (\zeta ^{+}) ~,~ 0 ~,~ 0 
~,~ 0 \right) ~,
\end{equation}
where $\zeta ^{+} = t + z$, in light front coordinates, {\it i.e.} $x^\mu = \{\zeta ^+, \zeta^ - , 1, 2 \}$
(the speed of light will be set to unity, $c = 1$). The $(+)$ in the superscript labels
this as the light front form of the solution traveling in the negative $z$ direction. The second solution gives 
waves traveling in the positive $z$ direction. The second solution is only a function of the light front coordinate
$\zeta ^{-} = t - z$,
\begin{equation}
\label{minus-old-coleman}
A^{(-) a} _\mu = \left (0 ~,~ x f^a(\zeta^{-}) + y g^a (\zeta^{-}) + h^a (\zeta ^{-}) ~,~ 0 ~,~ 0 \right) ~.
\end{equation}
Again, the superscript $(-)$ indicates this is the light front form of the solution traveling in the positive $z$ direction. 
The ansatz functions, $f^a (\zeta^{\pm}), g^a (\zeta ^{\pm})$ and $h^a (\zeta ^{\pm})$, are functions of
$\zeta^{\pm} = t \pm z$ but are otherwise arbitrary. First, plugging $A^{(+) a} _\mu$ from 
\eqref{old-coleman-a} into  \eqref{field-tensor}, the field strength tensor for the light 
front form of the $(+)$ solution becomes
\begin{equation}
\label{coleman-f+}
F ^{(+) a} _{\mu \nu}  =\left(
\begin{array}{cccc}
0 & 0 & -f^a (\zeta ^{+}) &- g^a (\zeta ^{+})  \\
0 & 0 & 0 & 0 \\
f^a (\zeta ^{+}) & 0 & 0 & 0 \\
g^a (\zeta ^{+}) & 0 & 0 & 0 \\
\end{array} ~
\right) .
\end{equation}
The non-zero components here are $F ^{(+) a} _{+ 1} = -f^a (\zeta ^+ )$ and $F ^{(+)  a} _{+ 2} 
= -g^a (\zeta ^+ )$. Next plugging $A^{(-) a} _\mu$ from \eqref{minus-old-coleman} into  
\eqref{field-tensor}, the field strength tensor for the light front form of the $(-)$ solution becomes
\begin{equation}
\label{coleman-f-}
F ^{(-) a} _{\mu \nu}  =\left(
\begin{array}{cccc}
0 & 0 & 0 & 0  \\
0 & 0 & f^a (\zeta ^{-}) & g^a (\zeta ^{-})  \\
0 & -f^a (\zeta ^{-}) & 0 & 0 \\
0 &  -g^a (\zeta ^{-}) & 0 & 0 \\
\end{array} ~
\right) .
\end{equation}
The non-zero components here are $F ^{(-) a} _{- 1} = -f^a (\zeta ^- )$ and $F ^{(-) a} _{- 2} = 
-g^a (\zeta ^- )$. It is interesting to note that neither of the forms of $F_{\mu \nu}^a$ depend on 
the ansatz function $h^a (\zeta ^\pm )$. Also, the non-Abelian term, $f^{abc} A^b _\mu A^c _\nu$, is always zero for
both the $(+)$ and $(-)$ solutions. Thus, the solutions are ``very weakly" non-Abelian since this prototypical 
non-Abelian/non-linear term is absent.

Coleman noted that the $(+)$ solution given by \eqref{old-coleman-a}, in terms of the vector potential, and by
\eqref{coleman-f+}, in terms of the fields, provides an example of the Wu-Yang ambiguity 
\cite{wu-yang-a} -- that in non-Abelian  theories, $F_{\mu \nu}^a$ does not contain all the 
gauge invariant information about a particular solution, as is the case in Abelian gauge theories. 
Specifically, in terms of the vector potential, one has the quantity 
\begin{equation}
\label{a-loop}
{\rm Tr} \left[ {\rm P} \exp \left( i \oint A_\mu ^a T^a dx^\mu \right) \right] ~,
\end{equation}
where ${\rm P}$ indicates path ordering and the $T^a$ are the Lie algebra elements. The expression in
\eqref{a-loop} is identified as the non-Abelian Aharonov-Bohm phase factor in terms of the potentials. As well, the expression in
\eqref{a-loop} is the Wilson loop for gauge theories \cite{wilson}. 

\begin{figure}
  \centering
	\includegraphics[trim = 0mm 0mm 0mm 0mm, clip, width=6.0cm]{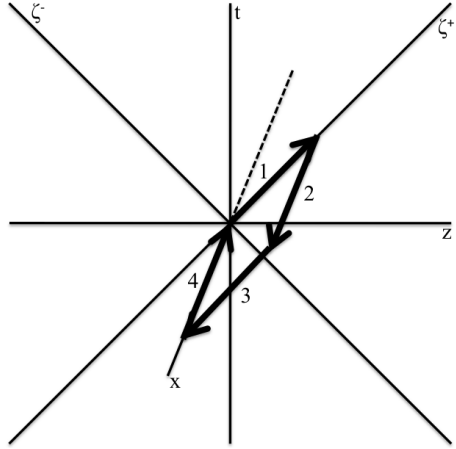}
\caption{{\it Unit loop in the $\zeta^+ - x$ plane}}
\label{fig1}
\end{figure}

We now consider a unit loop in the $\zeta^+ - x^1$ plane ({\it i.e.} $\zeta^+ - x$ plane) and starting from $x=0$ and
$\zeta^+ =0$ and going in the direction given in figure \ref{fig1}. 
For this loop the integral in the exponent in \eqref{a-loop} becomes 
\begin{eqnarray}
\label{a-loop1}
\oint A ^{(+) a} _\mu dx^\mu  &=& \int _1 A^{(+) a} _+ d \zeta ^+ + \int _2 A^{(+) a} _x dx  - \int _3 A^{(+) a} _+ d \zeta ^+ - \int _4 A^{(+) a} _x dx \nonumber \\
&=&  \int _1 h^a (\zeta ^+ ) d \zeta ^+ - \int _3 [ f^a (\zeta ^+ )  + h^a (\zeta ^+ ) ]  d \zeta ^+ ~.
\end{eqnarray}
The second and fourth integrals ({\it i.e.} $\int _2 A^{(+) a} _x dx$ and $\int _4 A^{(+) a} _x dx$) are zero since $A ^{(+) a} _x =0$ for
the $(+)$ solution in \eqref{old-coleman-a}. For the third integral, we get 
$- \int _3 [ f^a (\zeta ^+ )  + h^a (\zeta ^+ ) ] d \zeta ^+$, 
since for this leg, $x=1$ and $y=0$ and one is going backward along the $\zeta ^+$ direction, while for the first integral, we 
get $\int _1 h^a (\zeta ^+ )  d \zeta ^+$ since $x=y=0$. Note that $\int _1 h^a (\zeta ^+ )  d \zeta ^+$
and $- \int _3 h^a (\zeta ^+ )  d \zeta ^+$ do not cancel, due to the path ordering in \eqref{a-loop}. Taking the path ordering
into account and combining \eqref{a-loop1} and \eqref{a-loop} one gets
\begin{eqnarray}
\label{a-loop2}
{\rm Tr} \left[\exp \left( i T^a \int _1 h^a (\zeta ^+ ) d \zeta ^+ \right)
\exp \left( -  i T^b \int _3 [ f^b (\zeta ^+ )  + h^b (\zeta ^+ ) ]  d \zeta ^+ \right) \right] ~.
\end{eqnarray}
From \eqref{a-loop2} it is evident that there is no cancellation of the $h^a$ functions due to the non-trivial commutation relationship of
$T^a$ and $T^b$. The above result is equivalent to the result given in equation (8) of \cite{coleman}.

Now, the field strength version of \eqref{a-loop} (and the field strength version of the Aharonov-Bohm phase for 
non-Abelian theories \cite{peskin}) is
\begin{equation} 
\label{f-loop}
{\rm Tr} \left[ {\cal P} \exp \left( \frac{i}{2} \int F_{\mu \nu} ^a T^a d \sigma ^{\mu \nu} \right) \right] ~,
\end{equation}
where $d \sigma ^{\mu \nu}$ is the area and ${\cal P}$ means ``area" ordering \cite{borda}. For the unit area spanning the
unit loop in the $\zeta^+ - x$ plane, the differential ``area" is given by 
$d \sigma ^{+ 1} = d\zeta^- ~dx = (dt - dz) dx$. The reason that
$d \sigma ^{+ 1}$ has $d\zeta^-$ rather than $d\zeta^+$ is that the ``area" vector should be perpendicular
to the surface spanned by $\zeta^+ - x$, and it is $\zeta^-$, not $\zeta^+$, which is perpendicular to
$d \sigma ^{+ 1}$, as seen in figure \ref{fig1}. Similarly, $\zeta ^+$ is perpendicular to $d \sigma ^{- 1}$.
Note that for this unit square the area ordering denoted by ${\cal P}$ is simple since there is only one area vector.
With only a single area vector the issue of ordering does not arise. As a result of the above discussion, the integral in
the exponential in \eqref{f-loop} for the unit loop in the $\zeta^+ - x^1$ plane becomes
\begin{equation}
\label{f-loop1}
\int F ^{(+) a} _{+ 1} T^a d \sigma ^{+ 1} = - T^a \int \left( f^a (t+z) \right) (dt ~ dx - dz ~ dx) = 0 ~.
\end{equation}
This integral is zero since $\int f^a (t+z) dt = \int f^a (t+z) dz$, which is due to the $\zeta ^+$ functional 
dependence of $f^a$, but there is a sign difference between the $dt$ integration and $dz$ integration. For a unit 
loop in the $\zeta^+ - x^2$ plane ({\it i.e.} $\zeta^+ - y$ plane), we would find a similar result as 
in \eqref{f-loop1}, except for the replacement $F ^{(+) a} _{+ 1} \rightarrow F ^{(+) a} _{+ 2} = - g^a (t+z)$. 
Since the area for this loop is $d \sigma ^{+ 2} = d\zeta^- ~dy = (dt - dz) ~ dy$, we again get 
$\int g^a (t+z) dt = \int g^a (t+z) dz$, which then cancels because of the $d \zeta ^-$ in the unit area
element. The same result also holds for the $(-)$ solution from \eqref{coleman-f-}. In this case, the unit loop
is in the $\zeta^- - x^1$ plane ({\it i.e.} $\zeta^- - x$ plane) or $\zeta^- - x^2$ plane 
({\it i.e.} $\zeta^- - y$ plane). The perpendicular areas in this case will be 
$d \sigma ^{- 1} = d\zeta^+ ~dx = (dt + dz) dx$ and $d \sigma ^{- 2} = d\zeta^+ ~dy = (dt + dz) ~ dy$. 
Now, the relevant integrals will be $\int f^a (t-z) dt = - \int f^a (t - z) dz$ and 
$\int g^a (t-z) dt = - \int g^a (t - z) dz$, so that one again gets zero for the area integrals like
$\int f^a (t-z) T^a (dt ~ dx + dz ~ dx) = 0$ and $\int g^a (t-z) T^a (dt ~ dy + dz ~ dy) = 0$. This vanishing of
the ``area" integral of the non-Abelian field strengths, $F ^{(\pm) a} _{\mu \nu}$, occurs 
for both of these time-dependent, non-Abelian solutions we examined. Although we do not have a general proof,
we conjecture that this cancellation will occur generally for time-dependent, non-Abelian solutions. 
An important point to note is that it is only the time-dependent part of  $\frac{1}{2} \int F_{\mu \nu} ^a T^a d \sigma ^{\mu \nu}$ 
which is conjectured to vanish -- the static parts of the fields still give the usual non-Abelian Aharonov-Bohm phase.

In the light front coordinates, the split between the non-Abelian electric and magnetic field components is not 
so straight forward. Just as the $\zeta ^\pm$ coordinates are mixtures of space and time coordinates, 
so too $F_{\pm \mu}^a$ are mixtures of the non-Abelian electric and magnetic field components. 
For this reason, we now perform the above analysis of $\oint A_\mu ^a T^a dx^\mu$
and $\frac{1}{2}  \int F_{\mu \nu} ^a T^a d \sigma ^{\mu \nu}$ in Cartesian coordinates $(t,x,y,z)$. We do this for the $(+)$ solution
of \eqref{old-coleman-a} but the same analysis applies to the $(-)$ solution of \eqref{minus-old-coleman}. The $x,y$ components
of the $(+)$ solution are the same in light front and Cartesian coordinates. The time and $z$ components are obtained
via $A_0 ^{(+) a} = \frac{1}{2} \left( A_+ ^{(+) a} + A_- ^{(+) a} \right)$ and 
$A_z ^{(+) a} = \frac{1}{2} \left( A_+ ^{(+) a} - A_- ^{(+) a} \right)$, which is the same way one transforms between
$\zeta _{\pm}$ and $t, z$. Also, note that the superscript $(+)$ labels the solution while the subscripts label the $\pm$
components of this solution. Thus, in Cartesian coordinates, the $(+)$ solution is 
\begin{equation}
\label{old-coleman-cart}
A^{(+) a}_\mu = \frac{1}{2} \left (x f^a(\zeta^{+}) + y g^a (\zeta ^{+}) + h^a (\zeta ^{+}) ~,~ 0 ~,~ 0 
~,~ x f^a(\zeta^{+}) + y g^a (\zeta ^{+}) + h^a (\zeta ^{+}) \right) ~,
\end{equation}
where now $\mu = (0, 1, 2, 3)$ rather than $\mu = (+, -, 1, 2)$. The Cartesian field strength tensor following from
\eqref{old-coleman-cart} is
\begin{equation}
\label{coleman-f+-cart}
F ^{(+) a} _{\mu \nu}  = \frac{1}{2} \left(
\begin{array}{cccc}
0 & -f^a (\zeta ^{+}) & -g^a (\zeta ^{+}) & 0 \\
f^a (\zeta ^{+}) & 0 & 0 &  f^a (\zeta ^{+}) \\
g^a (\zeta ^{+}) & 0 & 0 &  g^a (\zeta ^{+}) \\
0 & - f^a (\zeta ^{+}) &  - g^a (\zeta ^{+}) & 0 \\
\end{array} ~
\right) ,
\end{equation}
where again the indices $\mu$, $ \nu$ are given by $\mu , \nu = (0, 1, 2, 3)$ rather than $\mu , \nu = (+, -, 1, 2)$.
As a check, it is easy to verify that \eqref{old-coleman-cart} and \eqref{coleman-f+-cart} satisfy the
Yang-Mills field equations \eqref{ym-field-eqn}. In the form \eqref{coleman-f+-cart}, the split between 
the non-Abelian electric and magnetic components is obvious -- the first row and column are the electric components and the 
$3 \times 3$ sub matrix below and to the right of the first row and first column are the magnetic components.  

Thus, in Cartesian coordinates, the loop integral in the exponent in \eqref{a-loop} becomes 
\begin{eqnarray}
\label{a-loop3}
\oint T^a A ^{(+) a} _\mu dx^\mu  &=& \frac{1}{2} 
\left( \int _1 T^a A^{(+) a} _0 dt + \int _1 T^a A^{(+) a} _z dz \right) 
- \frac{1}{2} \left( \int _3 T^b A^{(+) b} _0 dt + \int _3 T^b A^{(+) b} _z dz \right)  \nonumber \\
&=&  \int _1 T^a h^a (\zeta ^+ ) (dt ~ {\rm or} ~ dz) - \int _3 T^b [ f^b (\zeta ^+ )  + h^b (\zeta ^+ ) ]  (dt ~ {\rm or } ~ dz) ~.
\end{eqnarray}  
Again, the paths $2$ and $4$ do not contribute since the $x$-component of the $(+)$ solution in Cartesian coordinates is
zero, $A _x ^{(+)a} = 0$, as was the case for the solution in light front coordinates. Paths $1$ and $3$ have equal components in 
the $t$ and $z$ directions and thus, pick out the gauge field components $A _0 ^{(+)a}$ and $A _3 ^{(+)a}$. Along path $1$,
$x=y=0$, so only $h^a$ appears; while for path $3$, $x=1$ and $y=0$, so now both $h^a$ and $f^a$ appear. Because of the 
dependence of $h^a , f^a$ on $\zeta ^+ = t + z$, the $dt$ and $dz$ integrals of these functions are the same. The end result 
is that the loop integral of the gauge field gives the same results in light front coordinates, \eqref{a-loop1}, and in 
Cartesian coordinates, \eqref{a-loop3}. Thus the phase shift calulated by the loop integral of the gauge fields -- 
equation \eqref{a-loop2} -- is the same in both Cartesian and light front coordinates. 

Next, we calculate the surface area integral, $\frac{1}{2} \int F_{\mu \nu} ^a T^a d \sigma ^{\mu \nu}$,
for the unit area from figure \ref{fig1}. Taking into account the anti-symmetry of $F_{\mu \nu} ^a$ and 
$d \sigma ^{\mu \nu}$ under exchange of the indices, the $(+)$ solution, in Cartesian coordinates, and with the
surface in the $\zeta ^+ - x$ plane from figure \ref{fig1}, we find that the surface area integral of the
field strength tensor is
\begin{eqnarray}
\label{f-loop2}
\frac{1}{2} \int F ^{(+) a} _{\mu \nu} T^a d \sigma ^{\mu \nu} &=& T^a \left( \int  F_{01} ^{(+) a} dx^0  dx^1
+  \int  F_{13} ^{(+) a}  dx^1 dx^3 \right) \nonumber \\
&=&  T^a \left( \int  (- f^a (\zeta^+)) dt ~ dx +  \int  f^a (\zeta^+) dx ~ dz  \right)~.
\end{eqnarray}
There are two integrals -- $\int  F_{02} ^{(+) a} dx^0  dx^2$ and $\int  F_{23} ^{(+) a} dx^2 dx^3$ --
which are absent from \eqref{f-loop2} since for the unit area in the $\zeta ^+ - x$ plane, $dx^2 = dy=0$. If we take a surface in the
$\zeta ^+ - y$ plane, then these two integrals would appear and instead, $\int  F_{01} ^{(+) a} dx^0  dx^1$
and $\int  F_{13} ^{(+) a}  dx^1 dx^3$ from \eqref{f-loop2} would vanish. 
The two integrals in \eqref{f-loop2} cancel since $\int f^a (t+z) dt = \int f^a (t+z) dz$, due to the
fact that $f^a$ depends on $\zeta ^+ = t + z$. Similarly, the two integrals for the $\zeta^+ - y$ plane cancel since 
$\int g^a (t+z) dt = \int g^a (t+z) dz$, due to the fact that $g^a$ depends on $\zeta ^+ = t + z$. Thus, in both the
light front and Cartesian form, the $(+)$ solution has $\frac{1}{2} \int F ^{(+) a} _{\mu \nu} T^a d \sigma ^{\mu \nu}=0$.
In \eqref{f-loop2}, it is more apparent that the cancellation is between the non-Abelian electric pieces (the $F_{0i}^a$ integrals) and the corresponding non-Abelian magnetic pieces (the $F_{ij}^a$ integrals).  

As Coleman already noted, the expression $\oint A_\mu ^a T^a dx^\mu$ in \eqref{a-loop1} does not in general agree with 
$\int F_{\mu \nu} ^a T^a d \sigma ^{\mu \nu}$ in \eqref{f-loop1} for the light front solutions in 
\eqref{old-coleman-a} and \eqref{minus-old-coleman}. We have shown that the same is true (as it should be) in
Cartesian coordinate version of the solution. This is taken by Coleman \cite{coleman} as an illustration of 
the Wu-Yang ambiguity \cite{wu-yang-a} -- in non-Abelian gauge theories, not all the gauge invariant information 
is contained in the field strength tensor. 

\section{Time-dependent non-Abelian Wu-Yang monopole solution}
\label{t-depend-wy-mono}
We now examine the issue of the time-dependent, non-Abelian Aharonov-Bohm phase for the time-dependent $SU(2)$ Wu-Yang
monopole solution, where $f^{abc} \rightarrow \epsilon ^{abc}$. This solution is different in two respects from the non-Abelian plane-waves of the previous section.
First, in contrast to the non-Abelian plane wave solution of the previous section, the time-dependent $SU(2)$ Wu-Yang 
monopole solution is more strongly non-Abelian since the prototypical non-Abelian term 
($\epsilon ^{abc} A^b _\mu A^c _\nu$) is non-zero. Second, in the light front coordinates of the previous section
the non-Abelian electric and magnetic components of the field strength tensor are mixed up (combined) in non-zero
components like $F ^{(+) a} _{+ 1}$ or $F ^{(+) a} _{+ 2}$. For the time-dependent $SU(2)$ Wu-Yang monopole solution of this section,  
the field strength tensor is split into separate non-Abelian electric ({\it e.g.} $E_x ^a = F_{01} ^a$) and magnetic 
({\it e.g.} $B^a _z = - F_{12} ^a$) pieces. The vector potential for the time-dependent Wu-Yang solution is given by \cite{arodz}
\begin{equation}
\label{wy-time}
A_0 ^a = 0 ~~~~; ~~~~ A_i ^a = -\epsilon_{aij} \frac{x^j}{r^2} \left[1 + f(r,t) \right] ~,
\end{equation}
where $f(r,t)$ is a radial and time dependent function and $\epsilon_{ijk}$ is the SU(2) Levi-Civita 
structure constant of the group. For this time-dependent solution, one works specifically 
with the SU(2) group, whereas for the non-Abelian plane-wave solution of the last section, the
Lie group was arbitrary. As pointed out in the previous section, the non-Abelian part
of the field strength tensor did not play a big role in the Coleman solutions. For this reason, the exact nature of
the non-Abelian group was not so crucial for the Coleman plane wave solutions. From \eqref{wy-time}, one can 
immediately find the associated field strength tensor as
\begin{eqnarray}
\label{wy-f-time}
F_{0i}^a &=&  -\epsilon_{aij} \frac{x^j}{r^2} {\dot f}(r,t)  \nonumber \\
F_{ij}^a &=& -\left( \frac{1+f}{r^2} \right)' \frac{x^i x^k \epsilon_{ajk} -x^j x^k \epsilon_{aik}}{r}
-2\frac{1+f}{r^2} \epsilon_{aji} - (1+f)^2 \frac{x^a x^k}{r^4} \epsilon_{ijk} ~,
\end{eqnarray}
where the dot denotes a time derivative and the prime denotes a radial derivative.
The first two terms in $F_{ij}^a$ are the Abelian/pure curl part of the non-Abelian magnetic field ({\it i.e.}
$\partial _i A_j ^a - \partial _j A_i ^a = [\nabla \times {\bf A}^a] _k$). Also the non-Abelian electric field,
$E^a _ i = F_{0i}^a = -\partial_t A_i ^a$, is Abelian in character since the prototypical, non-Abelian
piece, $\epsilon^{abc} A^b _0 A^c _i$, is not present. One can thus show, generally, that there is a cancellation 
between the Abelian part of the non-Abelian electric field and the Abelian part of the non-Abelian  
magnetic field, $[\nabla \times {\bf A}^a] _i$. For this Abelian part of the fields from \eqref{wy-f-time}, 
the ``area" integral that appears in the field strength version of the Aharonov-Bohm phase is
\begin{eqnarray}
\label{abelian}
&& \int {\bf E}^a \cdot d{\bf x} ~ dt + \int {\bf B}^a _{(Abelian)} \cdot d{\bf a} = - \int \partial_t {\bf A}^a \cdot d{\bf x} ~ dt 
+ \int \nabla \times {\bf A}^a \cdot d{\bf a} \nonumber \\
&=& -\oint {\bf A}^a \cdot d{\bf x} + \oint {\bf A}^a \cdot d{\bf x} = 0 ~,
\end{eqnarray} 
where for the first, ``electric area" integral, we have done the time integration to get $-\oint {\bf A}^a \cdot d{\bf x}$;
for the second, magnetic area integral, we have used Stokes' theorem to get $\oint {\bf A}^a \cdot d{\bf x}$, which
then cancels the first, electric term. Note, for the first, electric term, one leg of the ``area" is a time piece. 

\begin{figure}
  \centering
	\includegraphics[trim = 0mm 0mm 0mm 0mm, clip, width=6.0cm]{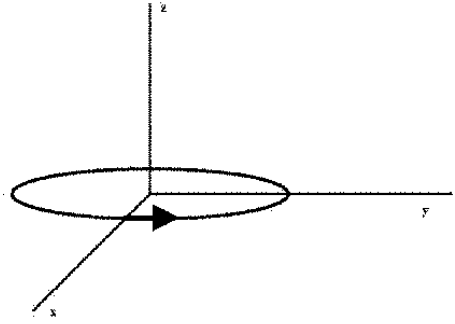}
\caption{{\it Hoop of radius $R$ in the x-y plane}}
\label{fig2}
\end{figure}

However, the last, non-Abelian term in the magnetic field -- {\it i.e.}  the term
$\epsilon_{ijk} \epsilon^{abc} A_i^b A_j^c = -(1+f)^2 \frac{x^a x^k}{r^4} \epsilon_{ijk}$ in \eqref{wy-f-time} --
could give a non-zero contribution to the calculation, so at first sight one would not, in general,
expect the same kind of cancellation between the non-Abelian electric
and magnetic parts that occurred in \eqref{f-loop1} for the Coleman solution. To this end, we will
look at the full non-Abelian fields of \eqref{wy-f-time} for a specific contour bounding a specific area, and we will
see that there is still a complete cancellation between the electric and magnetic parts, as there was for the Coleman 
solutions of the previous section. The specific contour we take is a hoop of radius $R$ in the $x-y$ plane going in the counterclockwise
direction. The area spanning this contour is a circle with area $\pi R^2 {\bf \hat z}$, {\it i.e.} pointing in the
$+{\bf \hat z}$ direction. The surface and contour are shown in figure \ref{fig2}. In evaluating the electric and
magnetic parts of \eqref{wy-f-time}, we will consider infinitesimal path lengths, $R \Delta \varphi$, and
infinitesimal areas, $R^2 \Delta \varphi / 2$ -- see figure \ref{fig3}.

We begin by looking at the non-Abelian electric contribution $\int F_{0i} ^a dx^0 dx^i = \int {\bf E}^a \cdot d{\bf x} ~ dt$. For the 
$F_{0i}^a$ from \eqref{wy-f-time}, one finds
\begin{equation}
\label{arodz-ef}
\int F_{0i} ^a dx^0 dx^i = -\epsilon_{aij} \int \frac{\dot f(R,t)}{R^2} x^j dt~ dx^i =
\frac{1}{R^2} \int [ {\bf x} \times d{\bf x}]^a \int \dot f(R,t) dt = \Delta \varphi \Delta f \delta^{a3} ~, 
\end{equation} 
where we have used the fact that $\int [ {\bf x} \times d{\bf x}]^a = \Delta \varphi R^2 \delta^{a3}$,
{\it i.e.} this integral gives twice the area of an infinitesimal wedge from the surface in figure \ref{fig2}
(see also figure \ref{fig3}). The direction of the area is in the $3$ or $+{\hat {\bf z}}$ direction. 
From the $\epsilon$ symbol in \eqref{arodz-ef}, with $i, j = 1,2$, the color index is fixed as $a =3$. 
To make $a=3$ explicit, we have inserted $\delta^{a3}$ into the final expression in
\eqref{arodz-ef}. The time integration has been done for an infinitesimal interval 
$\Delta t$, so $\dot f(R,t) \Delta t = \Delta f$.

Now we calculate the non-Abelian magnetic contribution to the phase. We do this in two separate pieces: the
Abelian part, $\partial _i A_j ^a - \partial _j A_i ^a $, and the non-Abelian part, 
$\epsilon_{ijk} \epsilon^{abc} A_i^b A_j^c$. First, for the Abelian part, we have
\begin{equation}
\label{abel-1}
B_{k ~ (Abelian)} ^a  = -\partial _i A_j ^a + \partial _j A_i ^a = \left( \frac{1+f}{r^2} \right)' \frac{x^i x^k 
\epsilon_{ajk} -x^j x^k \epsilon_{aik}}{r} +2\frac{1+f}{r^2} \epsilon_{aji} ~.
\end{equation}
For our contour from figure \ref{fig2}, the first term in \eqref{abel-1} has $i, j = 1, 2$ since we are in the $x-y$ plane.
But as well, for the summation over the $k$ index, we also need $k=1$ or $k=2$. For $k=3$, we would
have $x^3 = z$, but $z=0$ for the contour and surface we are using, so for $k=3$, the first term
in \eqref{abel-1} is zero. Taking all this into account, if we look at $i=1$ and $j=2$, we find that the 
indexed part of the first term in \eqref{abel-1} is 
$$
x^1 x^1 \epsilon _{a21} - x^2 x^2 \epsilon _{a12} = (- x^2 - y^2) \delta^{a3} =  -r^2 \delta^{a3} ~,
$$
since in the $x-y$ plane $x^2 + y^2 = r^2$. Note also that the color index, $a$, is forced to be $a=3$. With this, the
first term of \eqref{abel-1} becomes
\begin{equation}
\label{abel-3}
-\delta^{a3} \int \left( \frac{1+f}{r^2} \right)' r^2 dr d\varphi = 
-\delta^{a3} \int f' dr d \varphi + 2 \delta^{a3} \int \frac{(1+f)}{r} dr d \varphi~.
\end{equation}
Next, since our surface and contour from figure \ref{fig2} are in the $x-y$ plane, this means $i, j= 1, 2$, thus, for the last
term in \eqref{abel-1}, this implies that $a=3$, and we find that this term becomes
\begin{equation}
\label{abel-2}
2 \int \frac{(1+f)}{r^2} \epsilon_{321} dx^1 dx^2 = -2 \delta^{a3} \int \frac{(1+f)}{r} dr d\varphi ~,
\end{equation}
where in the $x-y$ plane $dx^1 dx^2$ become $r dr d \varphi$.

The second term in \eqref{abel-3} will cancel the term in \eqref{abel-2}. The first term in 
\eqref{abel-3} will have a $\Delta \varphi$ from the infinitesimal $\varphi$ integration. Then using infinitesimal 
notation for the $r$ integration ({\it i.e.} $\frac{\Delta f}{\Delta r} \Delta r = \Delta f$), we
arrive at 
\begin{equation}
\label{arodz-b-abel}
\int {\bf B}^a _{(Abelian)} \cdot d{\bf a} = -\Delta \varphi \Delta f \delta^{a3} ~,
\end{equation}
which then cancels the electric contribution \eqref{arodz-ef}. Thus, at this point we have confirmed, with specific 
contours and areas, the cancellation between the electric and ``Abelian magnetic" parts of the non-Abelian 
Aharonov-Bohm phase, which was shown generally in \eqref{abelian}. 

The final piece we need to deal with is the prototypical non-Abelian piece of the magnetic contribution, namely 
\begin{equation}
\label{arodz-a-na}
\int \epsilon_{ijk} \epsilon^{abc} A_i^b A_j^c dx^i dx^j = - \int (1+f)^2 \frac{x^a x^k}{r^4} \epsilon_{ijk} dx^i dx^j =0 ~.
\end{equation}
This piece is seen to vanish since $i, j = 1,2$ due to the contour/area from figure \ref{fig2} lying in the
$x-y$ plane. This forces $k=3$ so that $x^k \rightarrow z$, but since we are in the $x-y$ plane 
$z=0$, so this prototypical non-Abelian contribution vanishes. Thus, as for the Coleman plane wave solution of the
previous section, we find a cancellation between the electric and magnetic parts of the non-Abelian
Aharonov-Bohm phase. Although here we have shown this cancellation for only two types of time-dependent 
non-Abelian solutions and with specific contours, we nevertheless advance the hypothesis that this cancellation 
is a general feature of both Abelian and non-Abelian Aharonov-Bohm phases for time-dependent fields. 

We conclude this section by noting that, like the Coleman solutions, the time-dependent Wu-Yang monopole
solution shows a non-zero phase when calculated using the potential
\begin{equation}
\label{a-loop-arodz}
\oint A_\mu ^a dx^\mu \rightarrow -\oint {\bf A} \cdot d{\bf x} \rightarrow - \Delta A^a _i \Delta x ^i ~,
\end{equation}
where in the last step we are considering an infinitesimal path length as in figure \ref{fig3},
in conjunction with an infinitesimal change in the potential $\Delta A^a _i$. Using the form for $A^a _i$ from 
\eqref{wy-time}, we find
\begin{equation}
\label{a-loop-arodz1}
- \Delta A^a _i \Delta x ^i = - \Delta (1 + f(R, t)) \left( -\frac{1}{R^2}\epsilon_{aij} x^j \Delta x^i \right) 
\rightarrow - (\Delta f ) \left( \frac{1}{R^2} [{\bf x} \times \Delta {\bf x}]^a \right)~.
\end{equation}
In the last step, we have canceled two minus signs but have switched the $i$ and $j$ index which then gives
an additional minus sign. Also, we have used $\Delta f = \partial _t f(R, t) \Delta t = \frac{\Delta f}{\Delta t} \Delta t$.
Finally, we again use $[ {\bf x} \times \Delta {\bf x}]^a = \Delta \varphi R^2 \delta^{a3}$ and
find that
\begin{equation}
\label{a-loop-arodz2}
\oint A_\mu ^a dx^\mu \rightarrow -\oint {\bf A} \cdot d{\bf x} \rightarrow - \Delta A^a _i \Delta x ^i 
\rightarrow -\delta ^{a3} \Delta \varphi \Delta f ~.
\end{equation} 
This produces only the non-Abelian magnetic phase contribution from the fields calculation.
In the next section, we will discuss this non-equivalence between the time-dependent Aharonov-Bohm
phase shift, calculated using the potentials versus the field strengths, by comparing with the 
time-dependent {\it Abelian} Aharonov-Bohm phase case.

\section{Comparison with time-dependent Abelian Aharonov-Bohm effect}
\label{t-depend-abelian}
\begin{figure}
  \centering
	\includegraphics[trim = 0mm 0mm 0mm 0mm, clip, width=6.0cm]{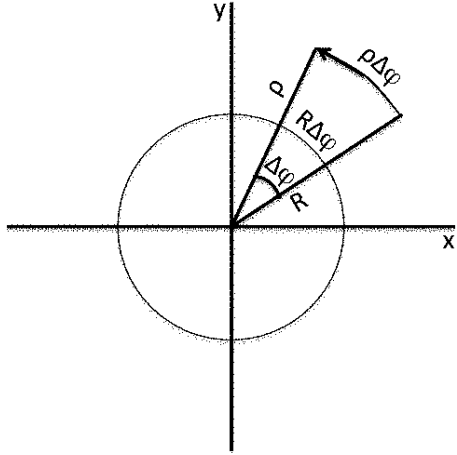}
\caption{{\it Infinitesimal path and area for the canonical Aharonov-Bohm set-up}}
\label{fig3}
\end{figure}

In the previous two sections, we discussed the time-dependent Aharonov-Bohm phase shift for non-Abelian fields for
two specific solutions (the Coleman plane wave solutions and the time-dependent Wu-Yang monopole) and two specific
contours (a unit square in the $\zeta^+ - x$ plane and a ring of radius $R$ in the $x-y$  plane). 
For the Coleman solutions, the field strength tensors had no contribution from the
non-Abelian term, $f^{abc} A_\mu^b A_\nu^c$, and the functional form of the field strengths was
essentially the same as for Abelian plane wave solutions. One distinction between the Coleman
non-Abelian plane waves and Abelian planes waves is that superposition does not apply
for the Coleman plane waves. For the time-dependent Wu-Yang monopole solution, the
electric field strength terms had no contribution from the non-Abelian part of the solution, but the magnetic field
did. However, this non-Abelian part for the magnetic field was found not to contribute to the Aharonov-Bohm phase
for the specific paths and surfaces we used, which are shown in figure \ref{fig2} (see also figure \ref{fig3}). We now review the Abelian, time-dependent Aharonov-Bohm effect and draw parallels with the non-Abelian case.    

The time-independent Abelian Aharonov-Bohm effect has been well studied theoretically and also been confirmed
experimentally \cite{chambers, tonomura} (see \cite{shikano} for a recent experimental tests using tunneling).
In contrast, the {\it time-dependent} Abelian Aharonov-Bohm effect has received much less attention. Some of 
the papers dealing with the time-dependent Abelian Aharonov-Bohm effect are \cite{chiao} \cite{kampen} 
\cite{singleton} \cite{singleton2}. The paper \cite{chiao} predicts a time-shifting interference pattern for 
the time-dependent Abelian Aharonov-Bohm effect, while \cite{singleton} \cite{singleton2} find
a cancellation of the time dependent electric and magnetic contributions and thus a non-shifting interference pattern.
The few experiments done on the time-dependent Abelian Aharonov-Bohm effect (see \cite{chentsov} \cite{ageev} and
also \cite{marton} for an ``accidental" test of the time-dependent Abelian Aharonov-Bohm effect) confirm that
predictions of \cite{singleton} \cite{singleton2}, but each experiment had problems. Thus, a definitive 
test of the time-dependent Abelian Aharonov-Bohm effect, to distinguish between the predictions of
\cite{chiao} and \cite{singleton} \cite{singleton2}, still needs to be done.

The vector potential for a time-dependent Aharonov-Bohm solenoid is given by \cite{singleton} \cite{singleton2}
(we use cylindrical coordinates $\rho, \varphi$ and the magnetic flux tube has a radius $R$)
\begin{eqnarray}
\label{3-vector-a}
{\bf A}_{{\rm in}} &=& \frac{\rho B(t)}{2} \hat{\bf \varphi} ~~~~{\rm for ~~ \rho <R} \nonumber \\
{\bf A}_{{\rm out}} &=& \frac{B(t) R^2}{2 \rho} \hat{\bf \varphi} ~~~~{\rm for ~~ \rho \ge R} ~.
\end{eqnarray}
To begin with, we have taken the scalar potential, $\phi$, as zero. We return to this point later since there are non-single 
valued gauges where there is a non-zero {\it and} non-single valued $\phi$. The possibility of a non-single valued  
$\phi$ leads to something similar to the Wu-Yang ambiguity but for time-dependent Abelian fields. 
The magnetic and electric fields coming from \eqref{3-vector-a} are
\begin{eqnarray}
\label{B-field}
{\bf B}_{{\rm in}} &=& \nabla \times {\bf A}_{{\rm in}} =  B(t) {\hat z}~~~~{\rm for ~~ \rho <R} \nonumber \\
{\bf B}_{{\rm out}} &=& \nabla \times {\bf A}_{{\rm out}} = 0  ~~~~{\rm for ~~ \rho \ge R} ~,
\end{eqnarray}
and 
\begin{eqnarray}
\label{E-field}
{\bf E}_{{\rm in}} &=& - \frac{\partial {\bf A}_{{\rm in}}}{\partial t} = - \frac{\rho {\dot B(t)}}{2} \hat{\bf \varphi}  
~~~~{\rm for ~~ \rho <R} \nonumber \\
{\bf E}_{{\rm out}} &=& - \frac{\partial {\bf A}_{{\rm out}}}{\partial t} =  
- \frac{{\dot B(t)} R^2}{2 \rho} \hat{\bf \varphi}
~~~~{\rm for ~~ \rho \ge R} ~.
\end{eqnarray}
Evaluating the Aharonov-Bohm phase using the fields \eqref{B-field} \eqref{E-field}, for the infinitesimal path and associated 
area in figure \ref{fig3} gives
\begin{eqnarray}
\label{ab-fields-phase}
\int {\bf E}_{\rm out} \cdot d{\bf x} dt + \int {\bf B}_{\rm in} \cdot d{\bf a} \rightarrow 
({\bf E}_{\rm out} \cdot \Delta {\bf x} \Delta t ) + ({\bf B}_{\rm in} \cdot \Delta {\bf a} ) ~.
\end{eqnarray}
By expanding ${\bf B}_{\rm in} = {\bf B}_0 + \dot{\bf B} \Delta t$ and identifying the infinitesimal path, 
$\Delta {\bf x} = \rho \Delta \varphi \hat{\bf{\varphi}}$ and the area, 
$\Delta {\bf a} = \frac{1}{2} R^2 \Delta \varphi \hat{\bf{z}}$, equation \eqref{ab-fields-phase} becomes
\begin{eqnarray}
\label{ab-fields-reduce}
- \left( \frac{\dot{B} R^2}{2} \Delta \varphi \Delta t \right) + \left( \frac{B_0 R^2 \Delta \varphi}{2} + 
\frac{\dot{B} R^2}{2} \Delta \varphi \Delta t \right) = \frac{B_0 R^2 \Delta \varphi}{2}~.
\end{eqnarray}
The time-dependent parts of the phase shift cancel each other, while the static 
Aharonov-Bohm phase shift, due to $B_0$, remains. 

Strictly, in order to extend the above analysis for an infinitesimal interval 
$\Delta t$ to an arbitrary time interval, one would need to consider a linearly
increasing magnetic flux, ${\bf B}_{\rm in} = {\bf B}_0 + {\bf B}_1 t$, where ${\bf B}_0$ and
${\bf B}_1$ are constant. For such a linearly changing magnetic field, all of Maxwell's equations 
are exactly satisfied with the fields given by \eqref{B-field} and \eqref{E-field}. 
However, for other time dependencies, the radial part of the vector potential will
be different. This arises due to the fact that for the form of the vector potential in
\eqref{3-vector-a}, the resulting electric and magnetic fields will not satisfy the Maxwell-Amp{\'e}re 
equation, $\nabla \times {\bf B} = \partial _t {\bf E}$, for finite time intervals.
As an example, for a sinusoidal time-dependence like $B(t) = B_0 e^{i \omega t}$, the vector 
potential will be given by Bessel functions \cite{singleton2} \cite{gaveau}
\begin{eqnarray}
\label{3-vector-a2}
{\bf A}_{{\rm in}} &=& A_1 J_1 (\omega \rho ) e^{i \omega t} \hat{\bf \varphi} ~~~~{\rm for ~~ \rho <R} \nonumber \\
{\bf A}_{{\rm out}} &=& \left[ C_1 J_1 (\omega \rho ) + D_1 Y_1 (\omega \rho ) \right] 
e^{i \omega t} \hat{\bf \varphi} ~~~~{\rm for ~~ \rho \ge R} ~,
\end{eqnarray}
where $A_1, C_1, D_1$ are constants and $J_1(x) , Y_1(x)$ are Bessel functions of order 1. For this
more complex form for the vector potential, one can still calculate the magnetic and electric fields as in
\eqref{B-field} and \eqref{E-field}; of course now ${\bf B}_{out} \ne 0$. However even with the
new form of the vector potential given in \eqref{3-vector-a2}, one can still make the same arguments
in \eqref{ab-fields-phase} and \eqref{ab-fields-reduce} which led to the cancellation of the time dependent
electric and magnetic contributions of the phase shift (see reference \cite{singleton2}). 
Thus one is left with only the static, time-independent contribution to the phase shift, given by the term 
$\frac{B_0 R^2 \Delta \varphi}{2}$ in \eqref{ab-fields-reduce}. If one evaluates the phase shift from the 
vector potential for the infinitesimal path in figure (\ref{fig3}) {\it at time} $t$, one obtains 
$\int {\bf A}_{\rm out} \cdot d{\bf x} \rightarrow \frac{B_0 R^2 \Delta \varphi}{2}$, which then matches the 
result in \eqref{ab-fields-reduce}.

Comparing the above calculations and discussion with the previous non-Abelian results, we see that
in both Abelian and non-Abelian theories there is a cancellation between the time-dependent 
electric and magnetic contributions to the Aharonov-Bohm phase shift. For the non-Abelian case
we have only shown this cancellation for two specific, time-dependent solutions and for special
contours and surfaces. Although we have not shown this cancellation for the non-Abelian fields 
in general, we nevertheless conjecture that this is a feature of more general time-dependent non-Abelian field
configurations. 

\section{Summary and Conclusions}

In this paper we have investigated the Aharonov-Bohm effect for {\it time-dependent} non-Abelian fields. In contrast to
the Abelian Aharonov-Bohm effect, much less work has been done on even the {\it time-independent} non-Abelian Aharonov-Bohm effect
-- reference \cite{Horvathy} is one of the few works to deal with the time-independent non-Abelian Aharonov-Bohm effect. The reason 
for this most likely lies in the difficulty in experimentally setting up and controlling non-Abelian field configurations. In comparison,
Abelian fields can be much more easily manipulated {\it e.g.} setting up the magnetic flux tube, which is used in the
canonical Abelian, Aharonov-Bohm setup. In this work, we studied (for the first time as far as we could determine from the
literature) the Aharonov-Bohm effect for time-dependent non-Abelian fields. We did this using two specific, known time-dependent 
solutions (the Coleman plane wave solutions \cite{coleman} and the time-dependent Wu-Yang monopole of \cite{arodz}) and using 
specific contours and associated areas (see figures \ref{fig1} and \ref{fig2}). There were two common results of this 
investigation: (i) The non-Abelian Aharonov-Bohm phase calculated via the
fields ({\it i.e.} $\int F_{\mu \nu} ^a T^a d \sigma ^{\mu \nu}$ see \eqref{f-loop}) did not agree in general with the
Aharonov-Bohm phase calculated via the potentials ({\it i.e.} $\oint A_\mu ^a T^a dx^\mu$ see \eqref{a-loop}). This point 
was already remarked on by Coleman \cite{coleman} as an example of the Wu-Yang ambiguity \cite{wu-yang-a} for non-Abelian 
fields. (ii) There was a cancellation of the time-dependent contribution to the Aharonov-Bohm phase shift coming from
the electric and magnetic non-Abelian fields . For the Coleman $(+)$ solution, this cancellation is given 
in equation \eqref{f-loop1} and for the time-dependent Wu-Yang monopole, this cancellation is given in equations 
\eqref{arodz-ef}, \eqref{arodz-b-abel} and \eqref{arodz-a-na}.

In section IV, we carried out a review of the time-dependent Abelian Aharonov-Bohm effect and made comparison to 
the results from the time-dependent non-Abelian Aharonov-Bohm effect from sections I-III. For both Abelian and 
non-Abelian fields, we found a cancellation of the time-dependent part of the phase shift. For the Abelian case, 
this was shown to occur generally -- see equations \eqref{ab-fields-phase} and \eqref{ab-fields-reduce}. This led 
to the conjecture that this cancellation was also a feature of more general time-dependent non-Abelian fields and 
more general contours/area. 

\section{Acknowledgments} M. B. was supported through a CSU Fresno College of Science and Mathematics FSSR grant 
and a Norma T. Craig grant.

%

%
%{\par\noindent {\bf Acknowledgments:}}
%
   
%

\end{document}